# Robust Myco-Composites: A Biocomposite Platform for Versatile Hybrid-Living Materials

Sabrina C. Shen[1a,b], Nicolas A. Lee[1a,c], William J. Lockett[a,d,e], Aliai D. Acuil[a,f], Hannah B. Gazdus[c,g], Branden N. Spitzer[a,b], and Markus J. Buehler[a,f,g,h]*

[1] Indicates equal contribution

[a] Laboratory for Atomistic and Molecular Mechanics (LAMM), Massachusetts Institute of Technology, 77 Massachusetts Ave. 1-165, Cambridge, MA, 02139, USA

[b] Department of Materials Science and Engineering, Massachusetts Institute of Technology, 77 Massachusetts Ave., Cambridge, MA, 02139, USA

[c] School of Architecture and Planning, Media Lab, Massachusetts Institute of Technology, 75 Amherst Street, Cambridge, MA, 02139, USA

[d] MIT Center for Art, Science & Technology (CAST), Massachusetts Institute of Technology, 77 Massachusetts Ave. 10-183, Cambridge, MA 02139, USA.

[e] Department of Media, Culture, and Communication, New York University, 239 Greene Street, New York, NY, 10003, USA

[f] Department of Civil and Environmental Engineering, Massachusetts Institute of Technology, 77 Massachusetts Ave., Cambridge, MA, 02139, USA

[g] Department of Mechanical Engineering, Massachusetts Institute of Technology, 77 Massachusetts Ave., Cambridge, MA, 02139, USA

[h] Center for Computational Science and Engineering, Schwarzman College of Computing, 77 Massachusetts Ave., Cambridge, MA, 02139, USA

* Corresponding author, mbuehler@MIT.EDU

**Abstract:** Fungal mycelium, a living network of filamentous threads, thrives on lignocellulosic waste and exhibits rapid growth, hydrophobicity, and intrinsic regeneration, offering a potential means to create next-generation sustainable and functional composites. However, existing hybrid-living mycelium composites (myco-composites) are tremendously constrained by conventional mold-based manufacturing processes, which are only compatible with simple geometries and coarse biomass substrates that enable gas exchange. Here we introduce a class of structural myco-composites manufactured with a novel platform that harnesses high-resolution biocomposite additive manufacturing and robust mycelium colonization with indirect inoculation. We leverage principles of hierarchical composite design and selective nutritional provision to create a robust myco-composite that is scalable, tunable, and compatible with complex geometries. To illustrate the versatility of this platform, we characterize the impact of mycelium colonization on mechanical and surface properties of the composite, finding that it yields the strongest mycelium composite reported to date, and demonstrate fabrication of unique foldable bio-welded containers and flexible mycelium textiles. This study bridges the gap between biocomposite and hybrid-living materials research, opening the door to advanced structural mycelium applications and demonstrating a novel platform for development of diverse hybrid-living materials.

# 1. Introduction

Living organisms have universally evolved the ability to transform abundant energy and materials from their environment into functional materials with extraordinary properties [1–4]. Recently, strategies to harness the abilities of various non-human organisms in materials engineering have received increased attention as a potential means of creating more sustainable materials with enhanced functionalities[5]. One such strategy is the creation of engineered hybrid-living materials, composed of live organisms grown on biotic or synthetic scaffolds, which enables engineers to harness material deposition and regenerative abilities of organisms while tuning factors such as form and substrate. This has led to innovative technologies including self-healing bacterial concrete[6], bacterial cellulose composites[7], and mycelium composites with proposed applications ranging from packaging to medical devices and construction materials[8,9].

Mycelium, the filamentous root structure of fungi, has particularly been investigated for its rapid growth rate and ability to upcycle cheap lignocellulosic waste (**Fig. 1a**). Mycelial hyphae quickly grow through loose substrates such as mulch or sawdust, binding them into solid, lightweight foams with potential as low-cost alternatives to synthetic packaging foams and insulators[10–14]. As a broader platform for sustainable materials, however, the scope of mycelium applications has been limited. This is due in large part to biological, engineering, and design constraints in conventional mold-based manufacturing methods where mycelium composites are formed by casting inoculated substrates within a prefabricated formwork[15–17]. Differences in humidity, oxygenation, and temperature throughout a mold can limit hyphal proliferation into the interior of larger forms, and the very nature of molded formworks limits geometries to those without enclosed voids or substantial overhangs. Oxygenation demands further require coarse substrates that allow oxygen diffusion throughout the mold, which precludes the creation of denser or more rigid composites as well as high-resolution material architectures. Finally, while some additive manufacturing systems for mycelium composites have been proposed, they generally suffer from limitations in scale and application due to the low stiffness of extruded material[18] or low resolution of printing resulting in the ability to only create crude geometries[19–21].

The upper limits of reported ultimate tensile strength and modulus of mycelium composites measure 0.20 MPa and 7.13 MPa respectively [14,22], falling within the range of lightweight foams such as expanded polystyrene foam (EPF) [16], but far weaker than conventional structural materials such as timber or particleboard[23]. Notably, some mycelium-based materials that require extensive post-processing such as heat pressing achieve good mechanical properties, but are outside of the scope of this work due to the required post-processing, which typically disables the living mycelium and limits fabrication to sheets and bricks[24,25]. In this work, we propose an approach to overcome the shortcomings associated with the manufacturing process and mechanical properties of conventional mycelium composites by utilizing a novel indirect inoculation method to induce fungal growth within mechanically robust 3D-printed biocomposites rather than bulky biomass residues (**Fig. 1b**). Importantly, this method allows for the realization of more complex geometries than mold-dependent methods and achieves great mechanical properties without requiring heat-pressing or extensive post processing. We use rational design and evaluation of key parameters in biocomposite formulation including component selection, hierarchically-induced anisotropy, and selective nutritional density to realize a rigid, printable, and biocompatible printing material, then characterize the surface and bulk impacts of mycelium colonization. The resulting mycelium composite achieves Young's modulus up to 160 MPa, over a 15-fold improvement compared to typical mycelium composites, to demonstrate the first 3D-printable mycelium composite with mechanical properties suitable for structural applications (**Fig. 1c, Table S1**). The platform described here elegantly leverages the strengths of biocomposite 3D printing technologies with high resolution, mechanical properties comparable to engineering materials, and capacity for creating complex geometries, while harnessing advantages of mycelium to achieve lower density, improved hydrophobicity, and generative capabilities. We further demonstrate the versatility of this method and the unique properties of mycelium materials that can be leveraged with the creation of self-sealing mycelium containers and flexible textile-like 3D prints and consider the broad extensibility of this platform to other applications and classes of engineered hybrid-living materials.

## 2. Results

### 2.1 Biocomposite Design

*2.1.1 Composition*

Several aspects of biocomposite formulation were considered in rational design to achieve a base material that was not only mechanically robust and compatible with fungal mycelium growth, but also amenable to high-resolution 3D-printing for fabrication of complex objects with diverse applications (**Fig. 2**). Importantly, the composite is fully composed of minimally processed natural materials derived from pre-existing waste streams, is entirely nontoxic and biodegradable, and can be printed in ambient conditions without heating elements or synthetic binding agents. Chitosan and cellulose were selected to form the composite's foundation due to their well-documented mechanical properties[26] and abundance in natural structural materials and agricultural waste[27]. Chitosan, the binding matrix, is a renewable byproduct of the shellfish industry considered to be biocompatible, biodegradable, and nontoxic, as well as a natural antimicrobial against some species of bacteria[28]. Cellulose is the most abundant biopolymer in the world and provides mechanical stability with its fibrous microstructure, similar to fiber-reinforced synthetic composites[29]. While a prior study suggested that a 1:8 chitosan to cellulose ratio yields pliable material[26], we found that a higher 1:2.5 ratio with lower hydration better maintained smooth printing and vertical layer integrity in high-resolution 3D-printing. This may be attributable to strong chitosan intermolecular forces[30], which can thicken the gel and enhance adhesion between printed layers. Decreasing water content minimized deformation and time spent in drying stages to a limit, where mixtures were too stiff to print.

To incentivize fungal growth on the composite, locally-sourced spent coffee grounds were incorporated to serve as a low-cost and sustainable source of nutrition[31]. As shown in **Fig. 2a**, with hydration level held constant, the addition of coffee increased biocompatibility with fungal mycelium as measured by hyphal growth rate. However, very high coffee concentrations eventually slowed mycelium colonization by incentivizing dense, localized growth[18]. High coffee concentration also reduced relative cellulose content, which negatively impacted vertical layer integrity during 3D-printing. This can be seen in **Vid. S1-3** and **Fig. S1**, which display the 3D-printing of chitosan-bound composites with only coffee, coffee and cellulose, and only cellulose respectively. Approximately 13% (w/v) coffee grounds was found to optimally balance robust mycelium growth and structural integrity during 3D printing. Pectin, a common fruit-derived carbohydrate that readily forms a gel in aqueous solution[32], was also considered as a binder and was hypothesized to enhance mycelial growth by providing nutrition without the negative structural effects of coffee grounds. While pectin-based composites showed robust initial mycelium growth, growth slowed dramatically within days and contamination by other species was frequently observed afterwards, indicating that competition with contaminant species for the highly nutritious pectin stalled fungal growth (**Fig. S2a**).

**Fig. 2, B and C** show the printed biocomposite along with the final composition that achieved excellent printing stability and robust mycelial growth. As a whole, key parameters for biocomposite design were found to be multiscale structural reinforcement, in this case involving intermolecular chitosan interactions and microscale cellulose fiber reinforcement, and appropriate nutritional density accessible only to the target organism. For instance, some fungi but few other organisms can subsist on spent coffee grounds, which are generally considered a poor nutritional source. An additional interesting consideration is that while coffee grounds negatively impacted structural stability, their porous nature may enhance oxygen diffusion, which is critical for mycelial proliferation[33,34]. Similar design principles can be applied to biocomposite substrates for other classes of engineered hybrid-living materials, such as with cellulose or mineral-depositing bacteria[35]. Here, the mycelium-optimized biocomposite was found to be compatible with several fungal species including the gourmet mushrooms Turkey Tail (*Trametes versicolor*) and Oyster (*Pleurotus ostreatus*)[36], and a strain engineered by Ecovative, a mycelium materials company[37] (**Fig. S2b**).

*2.1.2 Mechanical properties*

To first understand properties of the base composite without mycelium, bulk cast, fully infilled 3D-print, and 70% infilled 3D-print samples were fabricated with the optimized biocomposite in dogbone geometries for tensile testing, and walled cylinder geometries were 3D-printed for compression testing. Importantly, the behavior of 3D-printed samples depends heavily on their geometries, print parameters, and print toolpaths, so testing these samples does not produce data on absolute or bulk material properties, but provides relative information and perspective on what is achievable with this system.

The bulk mechanical properties of the composite varied slightly as compared to 3D-printed material, as summarized in **Fig. 2D**. Interestingly, in tension, the fully-infilled printed composite demonstrated an increased Young's modulus compared to cast samples, even with slightly lower density. This is likely attributable to the alignment of cellulose fibers, explored in detail below. Meanwhile, the modulus measured for printed material in compression was significantly reduced. This was expected as a consequence of the 3D printing toolpath since tensile samples were infilled at a 45º angle to the test vector, while compression samples were crushed perpendicularly to their vertical layers. This means that effects of mechanical defects such as imperfect infill and layer adhesion were much more pronounced in compression tests. Similar mechanical defects provide rationale for the printed composite's decreased tensile strength and elongation at break compared to bulk material in tension. Nevertheless, both the printed and cast composites achieve mechanical properties comparable to or exceeding current state-of-the-art biocomposites despite being optimized for biocompatibility and printability rather than solely mechanical properties. At approximate stiffness of 340 MPa and tensile strength of 4.2 MPa for the printed composite, this material is comparable to rigid polymer foams or a transverse section of wood[38]. Cast samples were not tested in compression due to difficulty fabricating high-quality compression samples in the same geometry as the printed compression samples, and because elastic modulus in tension and compression is theoretically the same in bulk materials. In applications where objectives other than biocompatibility or printability are paramount, similar biocomposites can be preferentially tuned such as by increasing the relative proportion of cellulose to enhance mechanical strength [26].

Evidently, sample geometry and print toolpath play a significant role in mechanical performance. This is analogous to hierarchical materials found in nature, where multiple levels of organization create extraordinary mechanical, optical, and other material properties[1–3]. Within this manufacturing system, the biocomposite was found to contain two notable levels of hierarchy, toolpath orientation and cellulose fiber alignment, which yield anisotropic mechanical properties based on toolpath orientation relative to the direction of applied strain (**Fig. 3A**). Toolpath orientation refers to the direction of the nozzle's movement relative to the vector of testing. Here, even in fully-infilled 3D prints, we observe that a 0º (parallel) toolpath orientation yields the highest observed tensile strength of 4.88 MPa, while a 90º (perpendicular) orientation yields the lowest at 1.45 MPa.

At the macroscale, toolpath orientation creates anisotropy from relatively weak adhesion between parallel paths. At the microscale, another contributor is the alignment of cellulose fibers, which straighten from shear stress as they are forced through a small extrusion nozzle, resulting in anisotropy within a single printed filament (**Fig. 3b**). **Fig. 3c** shows visible alignment of cellulose fibers in an SEM image of printed material, which can be quantified by calculating tangent divergence of fiber edges using equation 1 where $\phi_f$ is the 2D vector tangent measured on a fiber's edge and $\phi_p$ is the vector representing the print direction. This provides a metric of how aligned each fiber is with the print orientation, with a value of 1 or 0 indicating a parallel or perpendicular fiber respectively. Printed samples demonstrate a higher number of fibers aligned with the print direction. Tangent divergence was more evenly spread from 0-1 in the cast material, indicating a higher level of isotropy, however the small peak near 1 may have resulted from the scraping motion used to level cast material that can cause some surface alignment.

$$\text{Tangent Divergence} = \left\| \phi_f \cdot \phi_p \right\|$$
**(Equation 1)**

In this system, anisotropic effects from toolpath orientation and cellulose fiber alignment are coupled as a result of the extrusion method; that is, they are both aligned with the printing vector. Their joint effect results in printed composites that are strongest when strained along the axis of toolpath orientation and fiber alignment, weakest when strained in the perpendicular direction. These are important considerations in the design of printed objects and fabrication processes.

**2.2 Indirect inoculation**

Ecovative mycelium demonstrated faster growth and better contamination resistance than other species and so was used in subsequent experiments. Biocomposite samples were colonized with mycelium using an indirect inoculation method where printed samples were incubated atop a bed of living mycelium until fully colonized as indicated by development of a full mycelium skin around the sample, typically a period of 14 days (**Fig. 4a**). The removed samples could then be dried in order to stall fungal growth or kept hydrated in order to retain the behaviors of active mycelium.

Indirect inoculation provides a means of achieving rapid and robust mycelium colonization in high-resolution 3D printed forms. In sparsely infilled samples, void spaces were completely filled with mycelium hyphae (**Fig. 4b**), but even at fully dense infills (**Fig. 4c**), mycelium grew around and directly through the biocomposite to colonize it completely. It was empirically observed that mycelium grew readily along toolpaths, likely due to the ease of growing along a continuous filament with consistent nutrition, however the mycelium also spread across toolpaths quickly due to its ability to bridge small gaps[18]. The impact of toolpath geometry and complex forms on mycelium growth behavior, especially with large gaps, may have implications for toolpath design and warrants an independent future study.

Nevertheless, indirect inoculation has some inherent limitations. Because mycelium grows from an inoculation layer on only one side of the biocomposite print, despite rapid growth, full colonization theoretically takes longer than inoculating a print throughout as with direct inoculation methods where inoculum is mixed into composite material before printing. However, in our preliminary experiments, direct inoculation resulted in slow or stalled growth, likely due to the high shear stress experienced by mycelium when extruded through a thin nozzle, which is known to weaken living components[39]. Other studies have demonstrated successful mycelium colonization after printing with direct inoculation, however these generally use low resolutions and less rigid printing substrates[18,20,40] with poor mechanical properties[18]. As mycelium is a living material, much taller prints inoculated indirectly may exhibit inhomogeneity as mycelium closest to the inoculum ages or consumes more of its host material. When fabricating large-scale objects that do not require such high-resolution, this challenge can potentially be overcome by using direct inoculation methods with a larger nozzle size that reduces shear stress.

## 2.3 Impacts of mycelium colonization

### 2.3.1 Mechanical properties

Fully infilled, 70% infilled, and walled cylinder 3D-print geometries were fully colonized with indirect inoculation and dried prior to testing. The resulting mycelium composite was substantially denser than previously reported mycelium composites due to the use of a rigid 3D-printed substrate rather than loose biomass (**Fig. 1c,d**). At 160 MPa, our mycelium composites were also stronger than any other mycelium composite reported in literature, with over a fifteen-fold increase in modulus compared to the average of reported materials (**Fig. 4d**) excepting one by Kaiser et al., who similarly used a strong biocomposite substrate rather than loose particle biomass[41]. However, Kaiser et al. developed a thick and coarse "mycocrete" paste that was packed into molds rather than extrudable in high-resolution printing. In this work, tensile strength was measured at 0.72 MPa (**Fig. 4e**), similarly exceeding the highest performing mycelium composite in literature.

The mechanical properties of the mycelium-colonized composite are also shown in **Fig. 4d-f.** In comparison to the uncolonized biocomposite, growth of mycelium actually weakens the material. This is contrary to what is typically observed in particulate-based mycelium composites, where mycelium serves as a binder and enhances mechanical properties[22,42,43]. However, it makes sense in the context of a biocomposite substrate, where mycelium lends new properties but is actively digesting the organic material that makes up the pre-formed and already rigid material. This indicates a tradeoff between mycelium growth and mechanical properties in this class of materials, which presents an important consideration when selecting materials for specific applications. Kaiser et al. reported a tensile modulus of 153 MPa but did not report a tensile modulus of their base mycocrete due to slippage at clamps. They did report a base compressive modulus of 8.26 MPa, which was increased to 10.01 MPa after inoculation with mycelium. While this is notably weaker than our measured compressive modulus at 38.3 MPa, future work can investigate relationships between particulate-based and biocomposite-based substrates, and impacts of mycelium growth, to optimize for enhanced mechanical properties.

Interestingly, colonized samples of dense and sparse prints show relatively similar mechanical properties, especially compared to uncolonized samples of dense and sparse prints, indicating that the mechanical response in any colonized sample may be largely attributable to the mycelium. Additional detail regarding the mechanical behavior of the colonized myco-composite can be found in **Fig. S3**.

*2.3.2 Wettability and absorption*

Full colonization of biocomposites resulted in the formation of a smooth "skin" with improved hydrophobicity as measured by an increase in water contact angle from 117º to 138º (**Fig. 5a**). The mycelium skin further demonstrated minimal water absorption over time, with the contact angle reducing only 4.5º over 30 seconds (**Fig. 5b**). This may be attributable to the presence of hydrophobins, amphiphilic surface proteins produced by fungi to control surface interactions[44]. Over time, colonized material with the mycelium skin removed similarly showed an increase in hydrophobicity compared to uninoculated printed material, with the two materials demonstrating contact angle reductions of 29.0º and 52.1º respectively over 30 seconds. Uncolonized cast composite demonstrated a slightly smaller change in contact angle of 20.8º over the same period, likely resulting from its relatively higher density and therefore lower porosity.

In a similar vein, the fully colonized composite exhibited lower water absorption and volume swelling than non-colonized samples after being submerged for 24 hours despite initially swelling relatively rapidly (**Fig. 5c**), likely due to the presence of hydrophobic mycelial hyphae. The initial water absorption and swelling may be attributable to higher porosity. Submersion of materials with full mycelium skin intact was not investigated due to experimental challenges associated with growing samples with full mycelium skin on all surfaces, however we anticipate that such a coating would further reduce rates of water absorption and volume swelling[22].

*2.3.3 Bio-welded Containers and Textiles*
In preliminary experiments, mycelium was found to be capable of bridging gaps up to 8 mm between regions of printed biocomposite. To leverage this and other unique properties of mycelium with our additive manufacturing platform, two systems were designed to demonstrate novel capabilities. The first was a bio-welded mycelium container constructed by printing biocomposite panels separated by consistent gaps (**Fig. 6a**), then allowing mycelium to bridge these gaps to form flexible hinges along which the geometry could fold. After colonization, the panels were folded into an assembled box of living mycelium material, which continued to grow, sealing itself completely (**Fig. 6b**). This application uniquely leverages the bio-welding capabilities of mycelium to create a waterproof container that could, in theory, be opened and resealed as long as the fungi were kept alive and provided with sufficient nutritional media[45]. Given the known insulative and flame-resistant properties of mycelium[46,47], this method may provide an opportunity to produce resilient self-sealing, sustainable packaging systems. Custom packages could feasibly be printed to fit their contents with minimal waste and sealed autonomously, removing the need for adhesives during the packaging process.

Similarly, small islands of biocomposite were printed with precise spacing (**Fig. 6c**) and colonized to yield a flexible textile-like material that could bend and stretch in multiple directions (**Fig. 6d**). During flexing and stretching, biocomposite regions remained rigid while mycelium flexed. This process yields a simplified manner of constructing textile-like mycelium materials compared to existing mycelium-based leather alternatives that generally require intensive processes of cutting and laminating thin-sheets of pure mycelium[48]. Leveraging additive manufacturing and indirect inoculation further enables precise engineering of rigid and flexible regions in a textile, which determines how it moves, folds, and drapes[49,50]. Video documentation of sealed mycelium containers (**Vid. S4**) and flexible mycelium textiles (**Vid. S5**) can be found in the supplemental information.

## 3. Conclusion

We have described a rapid and reliable means to create robust myco-composites that require no external mold, no heat-pressing or extensive post processing, and achieve good hydrophobicity and mechanical properties superior to existing myco-composites. This platform leverages a biocomposite derived entirely from existing organic waste streams and designed for biocompatibility with fungal indirect inoculation through selective nutritional provision as well as excellent printability, which can enhance mechanical properties through cellulose fiber and toolpath alignment. Compatibility with additive manufacturing further enables broad versatility in applications, such as the creation of complex geometries, including gap-bridging geometries that leverage both the stiffness of the colonized myco-composite and the extended generative capabilities of mycelium alone. This provides a means to lend the strengths of biocomposites and mycelium materials to one another while countervailing their weaknesses, bridging the gap between the two research fields.

Mechanically strong myco-composites with water resistant exteriors could hold promise as advanced structural materials, while lightweight forms with 3D-printed sparse infills can leverage the insulative properties of mycelium as potential packaging systems[51], and flexible textile-like behaviors can be achieved by harnessing the generative behaviors of mycelium. In this work, we used Ecovative mycelium and achieved greater mechanical properties than any reported mycelium composites in literature, including other biocomposite-based materials[41] and Ecovative-based materials[52,53]. However, mycelium growth and characteristics are highly dependent on the species and strain as well as on environment. We present a successful platform for creating robust myco-composites, and future work with this platform utilizing other species of fungi, especially more widely available or "wild-type" species, would be beneficial.

Following the tremendous design space constituted by composite formulation, organism and species selection, and processing parameters, there is high potential for continued exploration of novel hybrid-living materials for further enhanced mechanical or other objective properties. With sufficient data, a broad range of previously unexplored applications may become feasible. Such research efforts typically require substantial wrought experimentation, however computation-assisted approaches to biomateriomics may accelerate the rate of material discovery or provide an avenue for more rapid optimization of specific properties[54,55]. This lays the foundation for future advances that could enable scalable, sustainable manufacturing with hybrid-living biocomposites.

## 4. Experimental Section/Methods

### 4.1 Mycelium cultivation

Malt extract agar plates were prepared for mycelium cultivation. 2 wt% malt extract (Briess Malt & Ingredients Co., WI) and 2 wt% agar (Spectrum Chemical, NJ) were dissolved in boiling filtered water, then sterilized in a 6-quart pressure cooker at 15 psi for 45 minutes. While warm, 20 mL each of agar solution was pipetted into 45 mm diameter petri dishes and allowed to cool completely before sealing with parafilm and refrigerated until use.

*P. ostreatus* (oyster mushroom) and *T. versicolor* (turkey tail) were purchased as liquid mycelium cultures from liquidfungi.com (Holiday, FL) and shipped to Cambridge, MA in insulated packaging. Ecovative culture was purchased as bulk solid inoculated hemp substrate (Grow-It-Yourself material) from Ecovative Design (Green Island, NY). 0.2 mL of liquid cultures or small pieces of solid inoculated material were used to propagate each species onto prepared agar plates. Petri dishes were fully colonized in approximately 1 week, after which they were refrigerated at 3ºC to preserve the living mycelium. Subsequent propagations of the mycelium were inoculated with prior-generation colonized agar dishes.

### 4.2 Composite design

To explore principles of composite design for biocompatibility and printability, parameters including coffee content, hydration level, and acidity were varied individually. To assess biocompatibility, for each material of interest, 25g of composite was pressed into a 45 mm diameter petri dish. A 1 cm diameter cylindrical core of material was removed from the center and replaced with a mycelium-colonized agar plug of the same size. Growth rate was then evaluated over time as measured by surface area coverage. These experiments were performed with replicates of 3 to assess variability.

To assess printability, each composite of interest was printed in the same geometry with significant overhang. Qualitative features such as vertical layer integrity, overhang stability, and print consistency were observed.

### 4.3 Composite mixing & printing

Used coffee grounds were obtained in bulk from local waste streams (Dunkin', Boston, MA) and dried under forced air for at least 72 hours prior to being powderized in a bladed grinder until particles were approximately 150 µm in diameter**.** Prior to composite mixing, coffee grounds were hydrated to 82% w/v (450g coffee:550g water) with filtered water and steam sterilized at 15 psi for 120 minutes.

Hydrated biocomposite pastes were mixed at 21º C in a sterile laboratory setting. For each batch of composite, 80 g of 85% deacetylated chitosan from Bulk Supplements (Henderson, NV) was combined with 990 ml of 21º C filtered water in a mechanical blender (Blendtec, Orem, UT) and pulsed once to homogenize the mixture. 10 ml of pure glacial acetic acid (VWR, Radnor, PA) was added to the mixture in order to form a gel. The mixture was then pulsed for 15 seconds. The resulting chitosan gel was transferred to an industrial mixer (Atosa, Westboro, MA) and combined with 200 g of 200 µm de-lignated cellulose fibers (Creafil, Chestertown, MD) on a medium speed for 2 minutes. The mixture was then scraped from the sides of the mixing bowl and mixed again on a high speed for 3 minutes to form a smooth and homogeneous paste. 350 g of hydrated coffee grounds was then combined into the cellulose-chitosan mixture and mixed on low for 30 seconds. The bowl was scraped down and then mixed again for 30 seconds on a medium setting to form a homogenous paste.

Preliminary experiments with pectin binders followed a similar protocol but combined 200g (20% w/v) pectin instead of chitosan with water before adding acetic acid. Experiments with both pectin and chitosan stacked the binders, combining both 20% w/v pectin and 8% w/v chitosan. Due to the nonlinear behavior of combinations of pectin and chitosan[56,57], these chitosan, pectin, and chitosan and pectin blended solutions empirically yielded similar viscosity solutions.

For additive manufacturing, the hydrated biocomposite paste was loaded into 300 ml cartridges and sealed with a polyethylene piston and 1 mm nozzle. Pastes were printed via pneumatic extrusion at 21º C and 30% humidity. Dogbone geometries were printed with 45º diagonal crosshatch infills. Extended dehydration processes were implemented to counteract reported issues of deformation during the drying process. Un-inoculated composites were printed onto an acrylic grid and placed on a wire rack covered with a tarp to ensure even dehydration across the print. After 24 hours, prints were transferred to a dehydrator and dried at 40º C for 18 hours. Dried prints were rested at 21º C and 30% humidity for at least 24 hours prior to mechanical testing in order to allow for some absorption of ambient humidity.

Cast composites were loaded into silicone molds and allowed to rest on the tarp covered wire rack for 24 hours of slow drying, then dehydrated at 40º C for 18 hours. Samples were removed from their silicone molds approximately 2 hours into the dehydration process. Dried samples were rested at 21º C and 30% humidity for at least 24 hours prior to mechanical testing.

### 4.4 Additive manufacturing system
The digital fabrication system consisted of a 3-axis gantry positioning system with a print area of 300 cm x 150 cm x 10 cm fitted with a pneumatic dispensing system (Nordson Medical[58], Salem, NH). The composite was typically extruded at 500 kPa with feed rate of 1250 mm/min. Extrusion pressure and feed rate were manually adjusted during printing to achieve consistent fabrication with 2.5 mm extrusion width, 2 mm layer height, and 0.5 mm system resolution.

### 4.5 Indirect inoculation
Purchased Ecovative substrate (inoculated hemp) was pressed flat into containers and incubated at room temperature for a period of five days prior to the insertion of 3D printed media. The remainder of the substrate was refrigerated at 3ºC until use.

Prints for indirect inoculation were extruded onto photopolymer 3D printed (Stratasys Inc, Revohot, Israel) grids. These were immediately transferred into airtight containers filled with Ecovative substrate and left to be colonized for 14 days prior to removal and dehydration at 40º C for 24 hours. Before mechanical testing, dried prints were allowed to rest for at least 24 hours at 21º C and 30% humidity.

### 4.6 Mechanical testing
Tensile testing was performed using an Instron Universal Testing Machine (Instron, Norwood, MA) with a 5 kN load cell. The ASTM d1037 dimensions for tensile testing were modified to use a gauge width of 24 mm in order to ensure more consistent breakage within the gauge. Test samples were strained at a rate of 1 mm/min until failure. Tests that exhibited breakage in the grips were discarded from statistical analysis.

Compression testing was performed using an Instron Universal Testing Machine (Instron, Norwood, MA) with a 100 kN load cell. The test samples were walled cylinders with 5 cm diameter and 1 cm wall thickness. Concentric circles were printed in order to achieve the 1 cm thickness. Tests were compressed at a rate of 2 mm/min.

For each parameter evaluated, 3-5 samples were included in final analysis.

### 4.7 Hydrophobicity, wettability, and absorbance

Hydrophobicity was measured through static advancing contact angle measurements, adapted from ASTM D7334. For each of cast, printed, colonized, and colonized without mycelium skin specimens, three 5 $\mu$L droplets of distilled water, dyed red with food coloring for visibility, were manually deposited on the surface with a micropipette under video recording. For each droplet, contact angle was measured at timesteps 0, 1, 10, and 30 seconds after deposition. Contact angle was measured with the Drop Shape Analysis plugin on ImageJ[59] (Bethesda, Maryland) 3 times each for each droplet and averaged.

Water absorbance and swelling experiments were adapted from ASTM D1037. At least 3 test specimens for each cast, printed, and printed and inoculated settings were cut into approximately 5cm by 5cm by 0.5cm squares, then sanded smooth with 60 grit sandpaper. They were dehydrated overnight at 40°C, then submerged under 1 inch of room temperature water. Volume and mass were measured for each sample prior to submersion, and after blotting at 2 hours and 24 hours of submersion.

### Author Contributions
Authors S.S. and N.L. contributed equally to this work.
Conceptualization: S.S., N.L., W.L., M.B.
Data collection: S.S., N.L., A.A., H.G., B.S.
Data curation and visualization: S.S., N.L., B.S.
Writing – original draft: S.S., N.L., W.L., A.A., B.S.
Writing – review & editing: S.S., N.L., M.B.

### Conflict of Interest
The authors declare no competing interests.

### Acknowledgements
We extend our thanks to Dr. James C. Weaver for his kind assistance with SEM imaging. We also extend our appreciation to the Dunkin' employees at the Stratton Student Center location (84 Massachusetts Ave) for their friendliness and efforts in collecting and donating spent coffee grounds.

# Figures

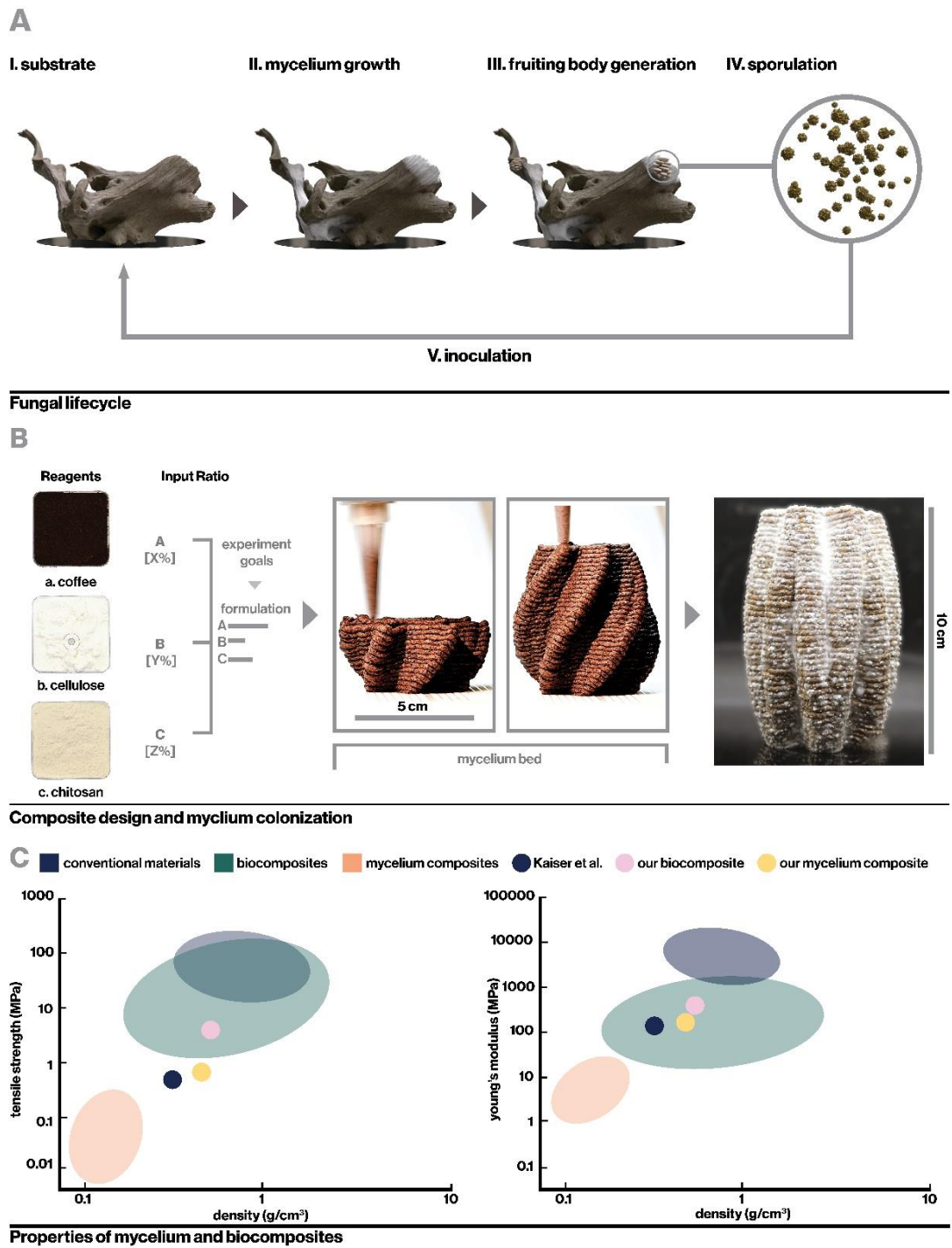

**Fig. 1. Platform for fabrication of functional and versatile engineered hybrid-living mycelium composites.** (**A**) Mycelium hyphae typically proliferate from spores in lignocellulosic media prior to the generation of fruiting bodies. (**B**) Overview of the design and fabrication process, including biocomposite design, additive manufacturing fabrication, and indirect inoculation of mycelium. (**C**) Mycelium materials generally demonstrate lower density, tensile and compressive strength, and Young's modulus as compared to both biocomposites and conventional materials. The described method of indirect inoculation in 3D printed biocomposites yields mycelium composites with improved strength and stiffness higher than all other reported mycelium composites in the literature to date. One mycelium-based material by Kaiser et al. achieves comparable mechanical properties by similarly utilizing a biocomposite base, but is not extrudable in high-resolution 3D printing.

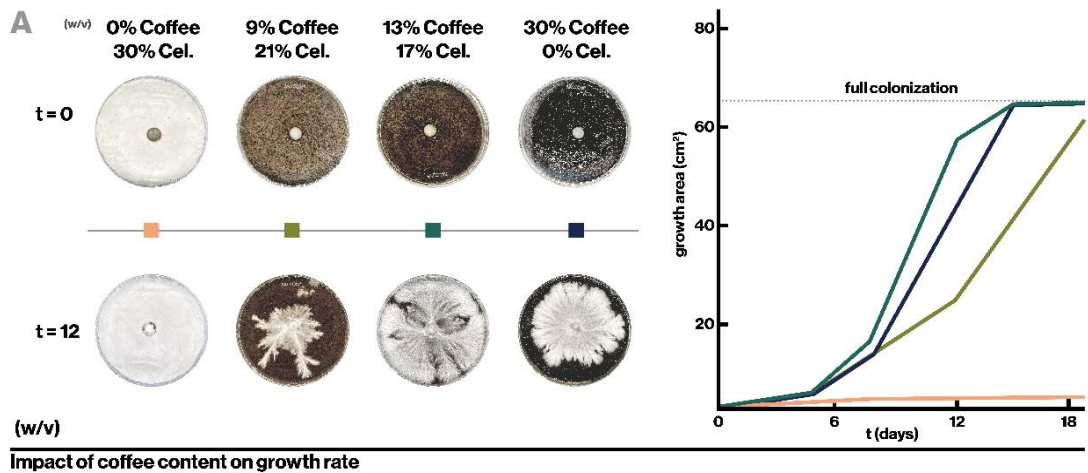
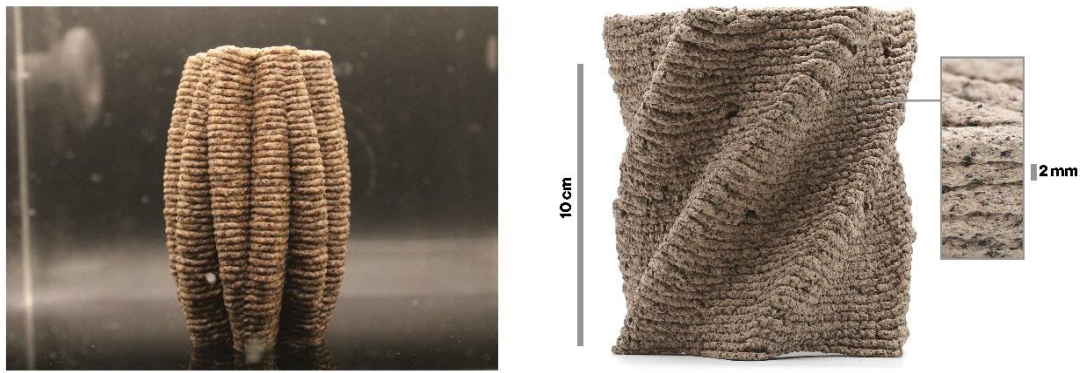
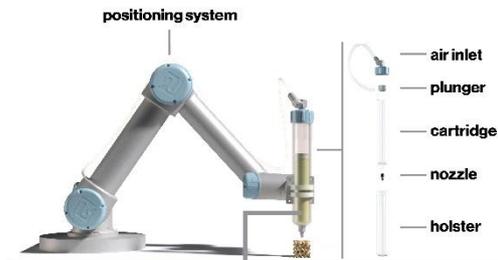
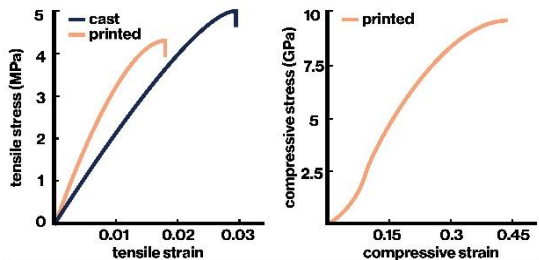

**Fig. 2. The biocomposite was optimized for mycelial compatibility and structural stability in printing and can be cast or additively manufactured into complex geometries while maintaining comparable mechanical properties.** (**A**) While holding the solid:liquid ratio constant, addition of spent coffee grounds enhanced mycelium colonization rate to a limit, where mycelium began to grow densely but slowly. Representative curves shown. (**B**) Hydrated (left) and solidified (right) biocomposites can be printed with a vertical resolution of 2 mm when hydrated. (**C**) A positioning system is affixed with a pneumatic extruder to extrude hydrated biocomposite at room temperature, which then solidifies via evaporation. (**D**) Even with decreased density, extruded composites yield higher Young's modulus than cast composites, potentially due to the alignment of cellulose fibers. However, extruded composites demonstrate decreased tensile and elongation at break, likely resulting from imperfect infill and inter-layer adhesion. Representative curves shown.

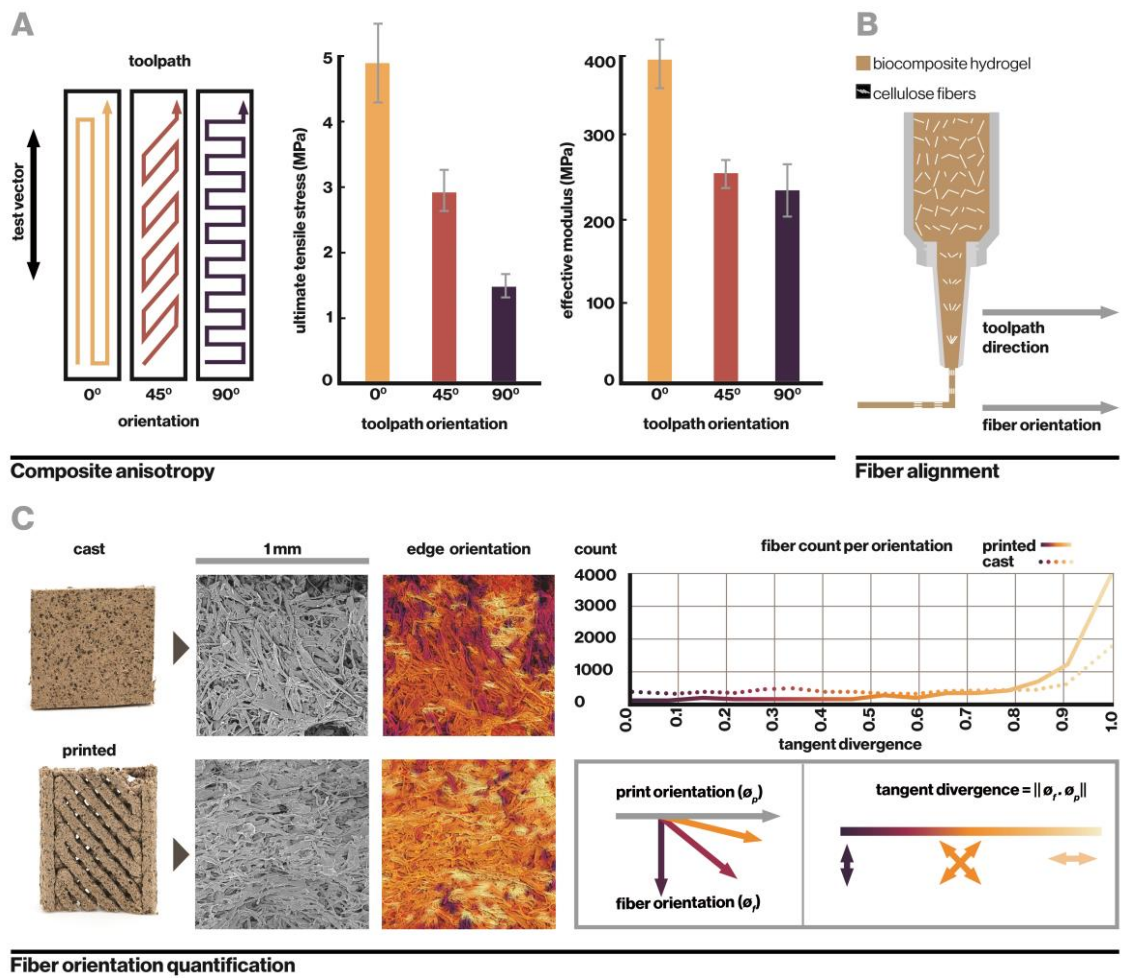

**Fig. 3. Additive manufacturing creates hierarchical anisotropy that impacts mechanical properties**. (**A**) Toolpath orientation creates macroscale printed filament alignment. This, along with microscale cellulose alignment, creates anisotropy in printed composites, with greatest ultimate tensile stress and Young's Modulus measured along the axis of print orientation. (**B**) Shear stress in the extrusion process forces microscale alignment of cellulose fibers in the same orientation as the toolpath direction. (**C**) Fiber alignment was quantified by visualizing cellulose fibers with SEM and measuring their dominant angle of alignment via the average tangent. The divergence of this tangent from the print orientation provides a quantification of whether fibers align with the toolpath. The printed sample shows high fiber alignment with the print orientation. The cast sample, which is expected to have randomly oriented fibers, shows a lower level of alignment that may result from manually spreading the material during the casting process.

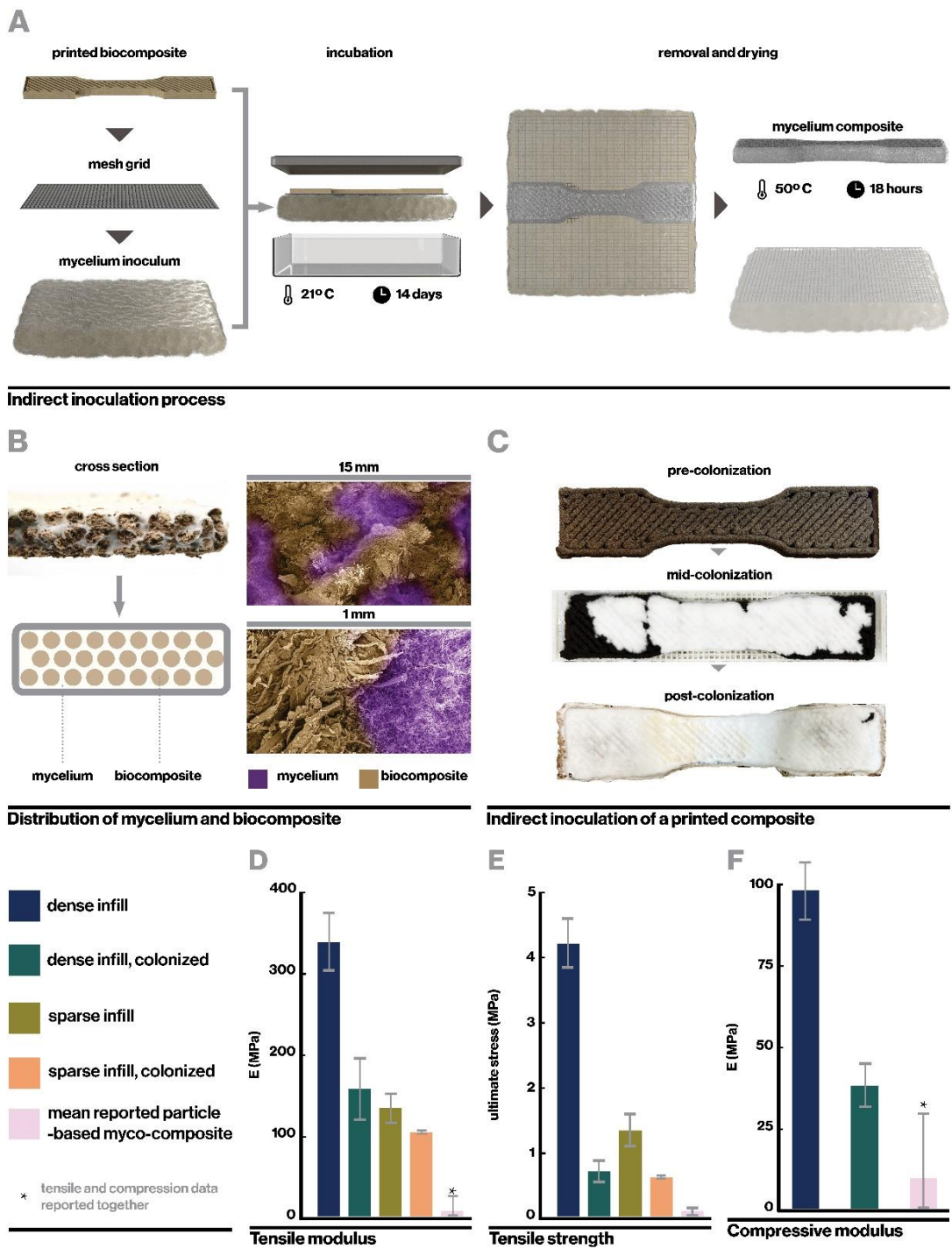

**Figure 4: Indirect inoculation allows for robust colonization of hydrated biocomposites, yielding a myco-composite with much improved mechanical properties relative to existing mycelium composites in literature.** (a) Immediately after printing, hydrated biocomposites are placed in contact with a bed of mycelium and sealed for incubation. Once full colonization has occurred, the colonized composite is removed and dried, yielding a solidified mycelium composite. (b) Cross sections of colonized 70% dense biocomposites show mycelial growth interspersed with biocomposite regions. Mycelium shows similar growth through 100% dense biocomposite prints, but does not form visually isolated pockets of mycelium. (c) Mycelium colonization growing through a 100% density infill in a 3D printed biocomposite. (d-f) Mycelium colonization decreases (d) tensile modulus, (e) tensile strength, and (f) compressive modulus relative to the uninoculated base composite due to digestion of some of the matrix materials. Nevertheless, the resulting myco-composites retain better mechanical properties than all myco-composites previously reported in literature, including a significant improvement over typical molded, particle-based myco-composites.

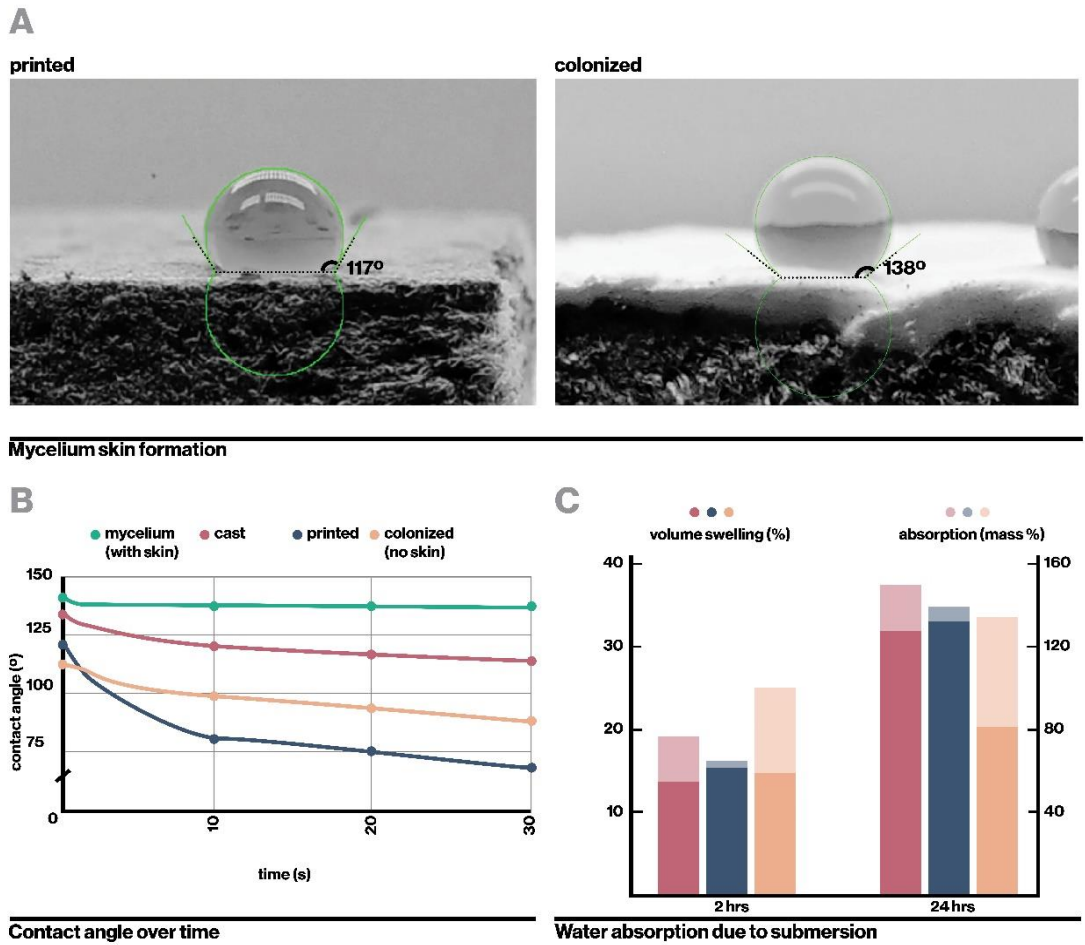

**Figure 5: In the described myco-composite, mycelium imbues hydrophobicity to its biocomposite substrate, especially with the mycelial skin intact.** (a) Droplet contact angle was measured over a period of 30 seconds to assess hydrophobicity. It was immediately apparent that mycelium skin demonstrates higher hydrophobicity than the base composite. (b) Mycelium skin demonstrates the most hydrophobicity and least water absorption over time. Colonized composite with the mycelial skin removed also demonstrates increased hydrophobicity compared to non-inoculated printed composite. (c) When submerged, colonized composites (without mycelium skin) demonstrate more rapid absorption of water initially, but ultimately absorb less and swell less than cast and printed composites. This may be attributable to its porous nature, which enhances absorption via capillary action, but the presence of hydrophobic mycelium, which then limits absorption.

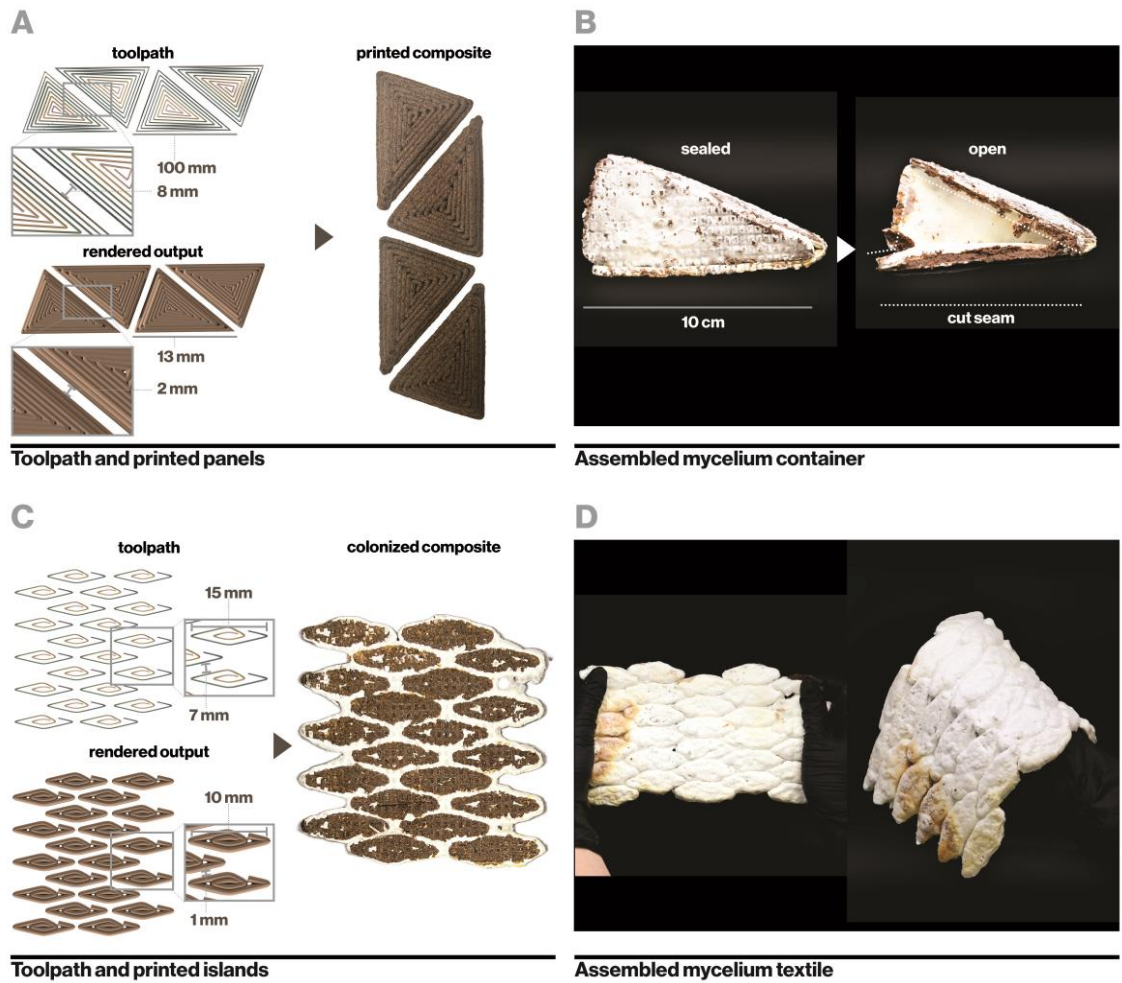

**Fig. 6. The described platform, additively manufacturing biocomposite geometries that are compatible with mycelium colonization via indirect inoculation, is highly versatile and enables utilization of mycelium-based materials in new applications**. (**A**) Disconnected panels were designed to have 2 mm gaps between pieces that would be bridged by mycelium during indirect inoculation. (**B**) The resulting "hinged" panels were assembled into a box by folding and mycelium growth was allowed to continue, sealing the seams with fungal material and yielding a fully-sealed container of mycelium biocomposite. (**C**) Similarly, small islands of rigid biocomposite can be printed adjacent to one another and flexible mycelium will bridge the gaps between them to form a flexible composite (**D**) that stretches, behaves like a textile, and can bend in multiple directions.

Supplementary Information

**Title: Robust Myco-Composites: A Biocomposite Platform for Versatile Hybrid-Living Materials**

*Sabrina C. Shen, Nic A. Lee, William J. Lockett, Aliai D. Acuil, Hannah B. Gazdus, Branden N. Spitzer, and Markus J. Buehler\**

| Material | Density (g/cm³) | UTS (MPa) | UCS (MPa) | Young's Modulus (MPa) | Reference |
|---|---|---|---|---|---|
| Douglas fir (softwood) panel | 0.44 | 130 | 50 | 11000 | 1,2 |
| Walnut (hardwood) panel | 0.55 | 89 | 52 | 11583 | 1,3 |
| (PLA) Poly(lactic acid) | 1.24 | 70 | 94 | 3120 | 4 |
| Plywood panel | 0.5 | 31 | 36 | 12400 | 5,6 |
| Medium density fiberboard | 0.78 | 18 | 10 | 3447 | 7 |
| ABS (Acrylonitrile butadiene styrene) | 1.06 | * | 49 | 2270 | 4,8 |
| Particle Board | 0.74 | * | * | 1999 | 9 |
| Chitosan film | 1.22 | 52 | * | 2200 | 10 |
| Corn starch film | 1.55 | 7 | * | 50 | 11 |
| Chitosan + cellulose | 0.37 | 6.12 | * | 263 | 12 |
| Cornstarch + corn husk | 1.3 | 13 | * | 325 | 11 |
| Chitosan + waste wood flour | 0.41 | 2.14 | 1.11 | 127 | 12 |
| Chitosan + wood flour | 0.31 | 1.63 | 1.05 | 97 | 12 |
| Cornstarch + corn husk/sugar palm | 1.3 | 17 | * | 1050 | 11 |
| Kenaf core fiber | 0.2 | * | * | 300 | 13 |
| Chitosan + silk fibroin | 1.46 | 119 | * | * | 14 |
| Ecovative + cotton fiber | 0.14 | 0.2 | * | 3.65 | 15 |
| Ecovative + hemp pith | 0.12 | 0.13 | 0.23 | 6.14 | 15 |
| Ecovative + hemp fiber | 0.1 | 0.1 | * | 7.13 | 15 |
| Pleurotus ostreatus + rapeseed straw | 0.13 | 0.01 | * | 2 | 16 |
| Pleurotus ostreatus + wheat | 0.18 | 0.05 | 0.04 | * | 17 |
| Trametes versicolor + rapeseed straw | 0.1 | 0.04 | * | 4 | 16 |
| Trametes versicolor + beech sawdust | 0.17 | 0.05 | * | 13 | 16 |
| Trametes versicolor + hemp | 0.13 | * | 0.36 | * | 18 |
| Trametes versicolor + loose hemp | 0.09 | * | * | 0.51 | 19 |
| Trametes versicolor + chopped hemp | 0.08 | * | * | 0.77 | 19 |
| Trametes versicolor + pre-compressed flax | 0.07 | * | * | 1.35 | 19 |
| Trametes versicolor + flax waste | 0.1 | * | * | 0.31 | 19 |
| Trametes versicolor + loose flax | 0.06 | * | * | 0.28 | [19] |
| Trametes versicolor + chopped flax | 0.07 | * | * | 1.18 | [19] |
| Trametes versicolor + wood | 0.09 | * | * | 0.14 | [19] |
| Trametes versicolor + wood chips | 0.17 | * | 0.52 | * | [18] |
| Irpex lacteus + mixed fiber media | 0.28 | * | 0.5 | 28 | [20] |
| Ganoderma lucidum + beech sawdust | 0.249 | 0.2 | * | 77.85 | [21] |
| Ganoderma lucidum + "mycrocrete" media | 0.306 | 0.52 | * | 153 | [21] |
| This work: Ecovative + chitosan/cellulose/coffee biocomposite | 0.466 | 0.72 | * | 160.27 | * |

**Table S1.**

Reported mechanical properties for conventional materials, biocomposites, and mycelium composites reported in literature. Mycelium composites typically demonstrate relatively low density, tensile strength, compressive strength, and Young's modulus.

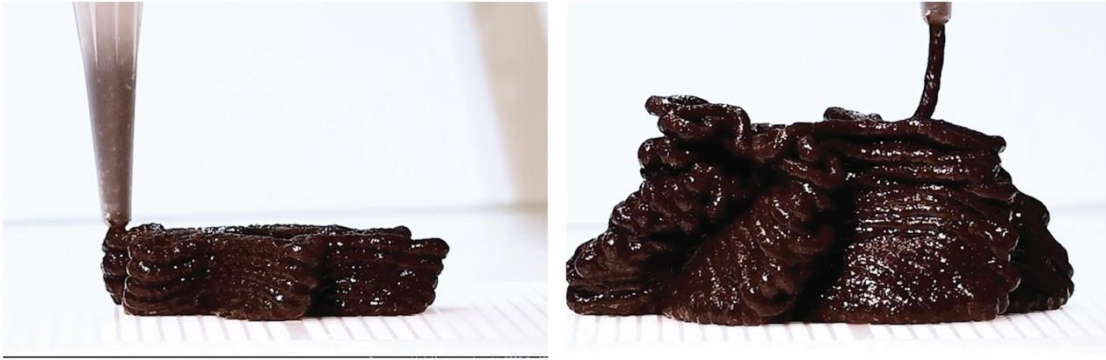
Additive manufacturing with 8% chitosan, 30% coffee, 0% cellulose (w/v)

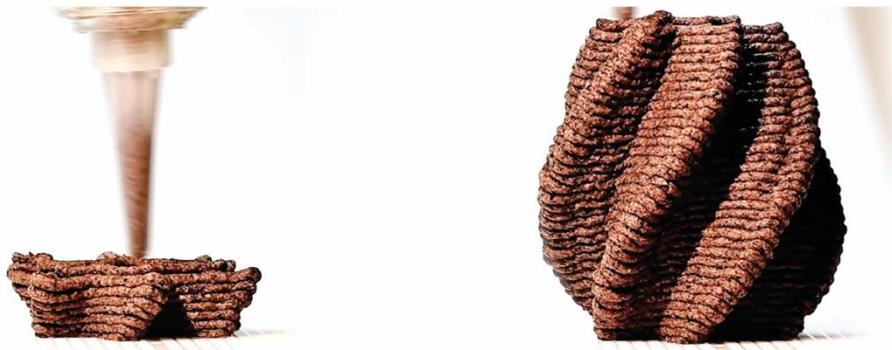
Additive manufacturing with 8% chitosan, 13% coffee, 17% cellulose (w/v)

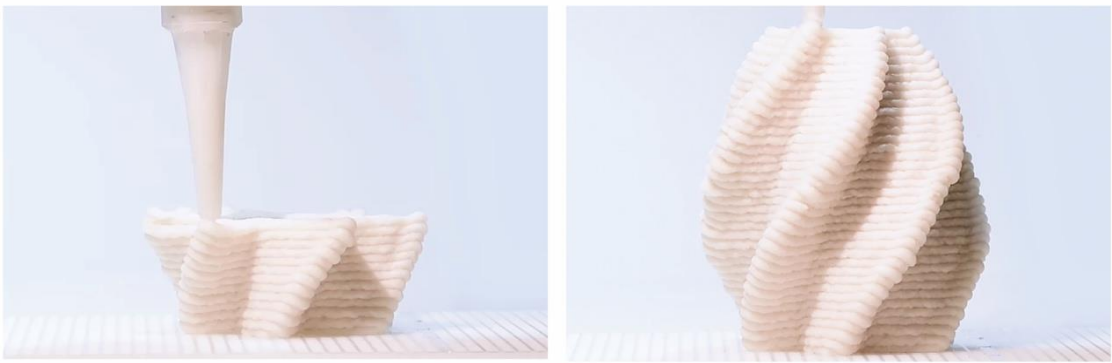
Additive manufacturing with 8% chitosan, 0% coffee, 30% cellulose (w/v)

**Fig. S1.**

The biocomposite composition was optimized for printability and biocompatibility with mycelium. (**A**) Chitosan-bound compositions containing high quantities of coffee but no cellulose did not maintain integrity while printing vertical layers. (**B**) An addition of 13% coffee grounds (w/v) provided a material with high integrity during vertical layering and fungal biocompatibility. (**C**) Composites with no coffee maintained high integrity during the printing process but did not encourage fungal growth.

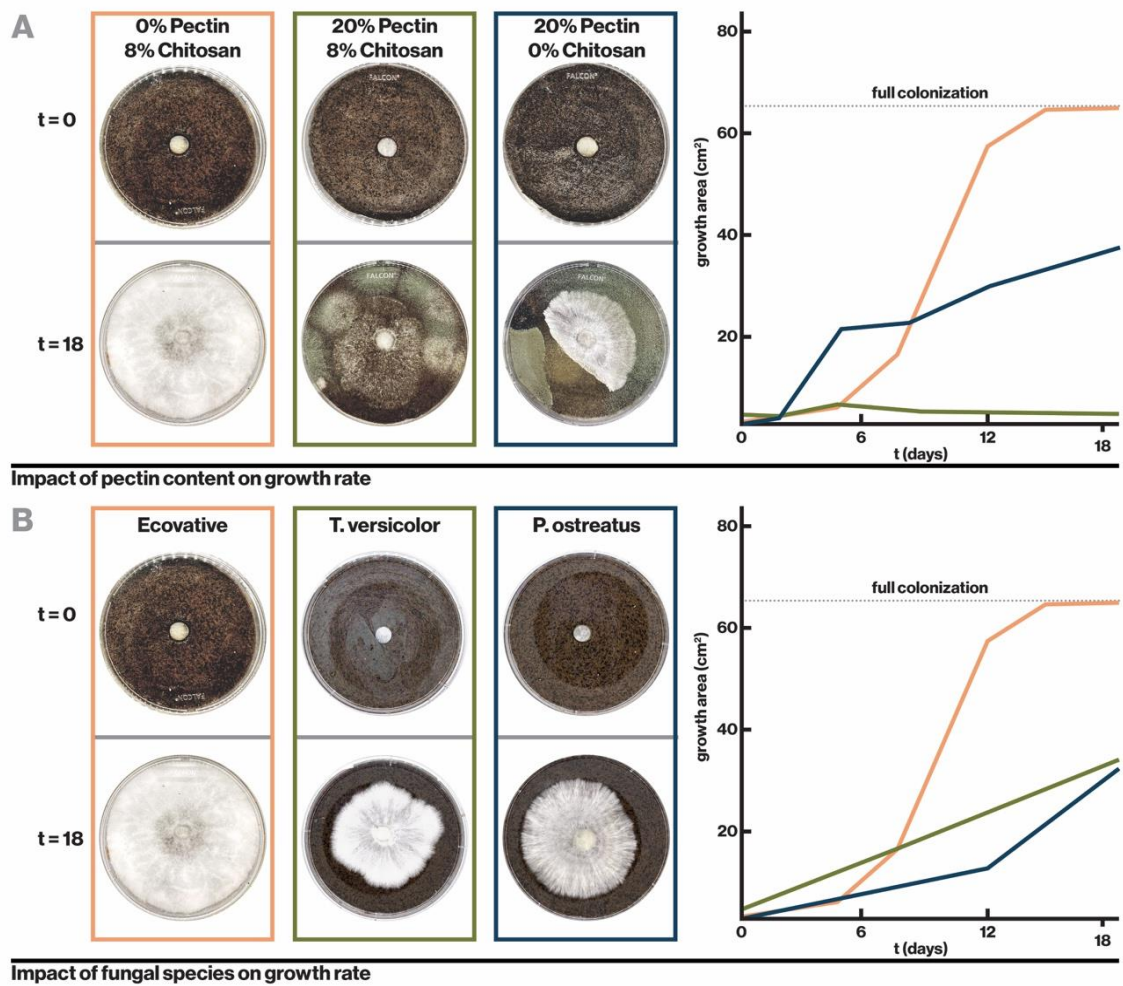

**Fig. S2.**

Impact of substrate pectin content and fungal species on growth rate. (**A**) Nutritious binding media containing pectin was explored as an alternative to coffee addition. Pectin-bound composites, which are rich in nutrition and calories, resulted in rapid initial growth followed by frequent contamination. This indicates that the nutrition source should be catered to the target organism. (**B**) Three species of fungi were explored using the optimized composite formulation. Ecovative mycelium was found to grow substantially faster than both T. versicolor and P. ostreatus strains on identical substrates, likely because it has been engineered for robust growth rather than gourmet fruit production.

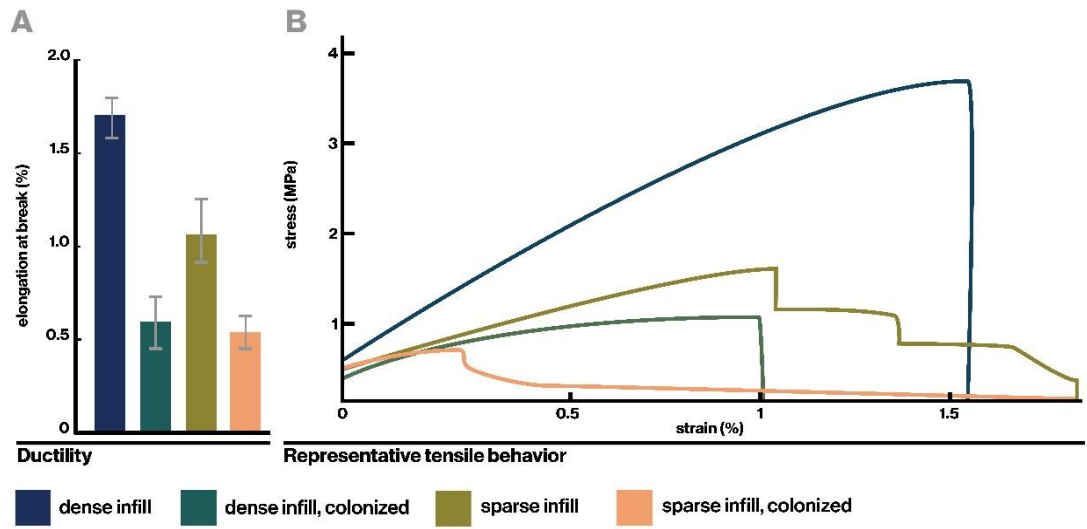

**Fig. S3.**

Colonization alters the tensile and compressive behavior of 3D printed samples with dense and sparse infills. (**A**) Elongation at break for dense and sparse samples, with and without colonization. Elongation is comparable for both dense and sparse samples when colonized with mycelium, indicating that mycelium behavior may be largely attributable for this response. (**B**) Representative curves for various samples. Tensile behavior is altered by colonization, with sparse-infill colonized samples exhibiting graceful failure rather than brittle fracture.

**Movie S1.**

Additive manufacturing of a chitosan-coffee composite, which shows limited structural integrity due to the lack of reinforcing structural components such as fibrous cellulose.

**Movie S2.**

Additive manufacturing of a chitosan-cellulose-coffee composite, which demonstrates good structural integrity.

**Movie S3.**

Additive manufacturing of a chitosan-cellulose composite, which demonstrates good structural integrity and smooth printability.

**Movie S4.**

A sealed container of living mycelium-biocomposite is cut open. The container is constructed from four flat panels of printed biocomposite spaced with 2 mm gaps between each panel. The mycelium grew to bridge the gaps, forming flexible hinges, at which point the box was folded into a 3D geometry. After a period of 7 days during which humidity was kept above 80%, mycelial growth proceeded to the point where the box was fully sealed and could be reopened using a knife. The void-space inside the container was maintained during growth.

**Movie S5.**

A flexible, living mycelium "textile" was constructed by printing islands of material with 1 mm spacing. Mycelium bridged the gaps to form flexible regions, allowing the thin sheet to be stretched or flexed in multiple directions without breaking.